# Analysis and particle-in-cell simulation of gridded ICRH plasma thruster

Ishaan Mishra[1]*

**Abstract:** Large-payload deep space missions are impractical with current rocket propulsion technologies in use. Chemical thrusters yield a high thrust but low efficiency while ion thrusters are efficient but provide too little thrust for large satellites and manned spacecraft. Plasma propulsion is a viable alternative with a higher thrust than electric ion thrusters and specific impulse far exceeding those of chemical rocket engines. In this paper, a hybrid thruster is explored which affords the high mass flow rate of plasma thrusters while maximizing the specific impulse. The two primary processes of this system are the ion cyclotron resonance heating of plasma and subsequent electrostatic acceleration of ions with gridded electrodes. Through a particle-in-cell simulation of these two components, the exhaust velocities of Xenon, Argon, and Helium are compared. It has been found that while the combination of systems results in a far greater exhaust velocity, the acceleration is largely from the gridded electrodes, and thus Xenon is the most suitable propellant with a specific impulse upward of 4200 s. Advancements in nuclear fusion and fission technologies will facilitate the deployment of high-power plasma thrusters that will enable spacecraft to travel farther and faster in the solar system.

[1]Physics and Optical Engineering, Rose-Hulman Institute of Technology, Terre Haute, IN, USA 47803
*Correspondence: mishrai@rose-hulman.edu

## 1. Introduction

As the 21st century opens up a new chapter in human space exploration, the methods of propulsion must also adapt to be more suitable for deep space missions. Current conventional chemical thruster technology that is employed for low-Earth orbit, Lunar, and Mars missions are not practical for deep space travel as the majority of the spacecrafts' mass is fuel due to the low thruster efficiency. While ion thrusters have proved to be highly efficient with a relatively long lifespan, the time required to complete missions is substantial for large payload missions beyond Earth's immediate environs. The long-term effects of such, continuous spaceflight by the increased radiation exposure and weightlessness are detrimental to the human body's functioning.

For the successful human expedition of space, propulsion systems are required which deliver a high thrust along with high efficiency. One solution is nuclear fission-driven rocket engines, in which energy released from fission reactions is used to accelerate propellants to provide thrust. However efficient containment of fissile chain reactions and cooling add to the existing issue with loading rockets up with highly radioactive materials. An alternate utilization of fusion in space propulsion is to power separate propulsion systems, typically those which require electric power [1].

Another technology that is relatively safer is nuclear fusion-driven propulsion, particularly since the primary fuel, Hydrogen and its isotopes, are found in abundance on both Jupiter and Saturn (10% and 25% respectively) [2, 3] which could be accessible in the future if technology for extraction is developed. As global efforts to advance nuclear fusion continue [4], the technology holds the key to revolutionising space propulsion, as a large source of power is available with relatively low fuel requirements, compared to conventional chemical thrusters. Fusion Technology can either be used to directly generate thrust, or to supply energy to high-power requiring propulsion systems, like plasma thrusters [5].

Over the last few decades, plasma propulsion has gained traction through projects like the VASIMR. In these systems, the propellant is ionized and subsequently accelerated after being heated to increase the exhaust velocity. These systems are advantageous to ion and chemical thrusters in that they afford both a high specific impulse and high thrust, and are thus best suited for interplanetary missions. In this paper, a thruster concept is explored that utilizes the gridded electrodes of ion propulsion within a plasma thruster in order to increase the thrust per unit mass of propellant by using electrostatic forces to further accelerate a heated plasma, in contrast to existing schemes which typically expel the plasma after heating and avoid electrode grids due to grid erosion that reduces the lifespan of the thruster. In recent years, progress has been made in developing electrode grids that minimize electrode grid erosion, with NASA's NEXT ion thruster that has a lifespan of 2.5 years [6]. Furthermore, this thruster has successfully operated in laboratory conditions for upward of 5 years [7]. It must be noted, however, that these long lifespans are facilitated by the low-thrust nature of ion thrusters, and the high-thrust concept that is explored in this paper would have a lower lifespan, which is apt as this model is intended for shorter-duration missions. A key assumption taken in this paper is the availability of a source of high electric power, which can be derived from fusion or fission reactors on the spacecraft.

## 2. Thruster Schematic

In order to maintain a high specific impulse along with a high thrust, the explored plasma thruster (Figure 1) consists of the following components:

- Plasma generation: Here, the propellant is converted into a plasma using helicon waves;
- Plasma Confinement & Heating: Cylindrical rings around the heating chamber are used to confine the plasma in a magnetic mirror. Here, the plasma is heated using Ion Cyclotron Resonance;
- Expansion through Nozzle: The plasma moves through a magnetic nozzle which converts the thermal energy of the heated plasma into kinetic energy along the thruster axis. Along the physical nozzle boundary walls are anodes that strip the plasma of its electrons. This is similar to the ionization chamber walls of ion thrusters.
- Gridded Electrostatic Acceleration: The plasma which now has a net positive charge is accelerated further using high voltage closely spaced gridded electrodes. The extracted electrons are removed from the spacecraft with a neutralizer.

*2.1 Plasma Generation*

For efficient plasma generation, a helicon wave antenna will be used which uses rf waves to ionize the neutral gaseous propellant. This is preferred over electron cyclotron resonance and other ionization methods as helicon plasma generation has been shown to have unusually high ionization efficiency, does not require internal electrodes or large sheath voltages, and offers a possible means of controlling the electron velocity distribution [8]. Furthermore, helicon antennas produce plasmas at a far higher density than electron cyclotron resonance and electrodes [9], where plasmas of densities exceeding $10^{13}$ cm$^{-3}$ can be obtained [10]. For this thruster, a double-saddle coil helicon antenna configuration is optimum for the plasma generation cylinder [11].

*2.2 Plasma Heating*

While efficient Helicon Thrusters have been tested and developed as alternatives to ion thrusters, the thrust imparted in these thrusters is low and is thus inadequate for missions with larger payloads [12, 13]. Thus, additional heating and acceleration are required for optimum thrust. An antenna heats up the plasma which is confined in a magnetic mirror [14].
Ion cyclotron resonance heating is a method of increasing the thermal energy of ions in magnetic fields using electromagnetic waves at the ion cyclotron resonance frequency, which can be calculated with the formula

$$f = \frac{qB}{2\pi m} \qquad (1)$$

where $f$ is the ion cyclotron resonance frequency, $q$ is the charge of the particle, $B$ is the magnetic field strength and $m$ is the mass of the particle. Electromagnetic waves at this frequency are absorbed by the ions in the magnetic field, thereby increasing their kinetic energy. This is a common heating method in tokamak reactors and has also been used in plasma thrusters like the VASIMR, where it results in a significant increase in the exit velocity of the propellant ions [15].

*2.3 Magnetic Nozzle*

To maximize the thrust from the ICRF-heated plasma, the propellant after heating undergoes acceleration through a magnetic nozzle. This is necessary as the increase in velocity of ions due to ICRH is not only along the direction of mass flow. As the plasma passes through the nozzle, particle temperature in the radial direction is converted to energy in the direction parallel to mass flow, and exit velocity increases by reducing the energy loss along the radial direction [16].

In this model, both a magnetic and physical nozzle is used as this combination has been shown to afford greater thrust than only a magnetic nozzle [17]. This physical nozzle is also necessary as electrodes will be placed here which strip the plasma of electrons. This is to prevent a charge build-up within the spacecraft that can impact performance as excess negative charges can interfere in the proper expulsion of the largely positive plasma (electron passage is restricted by the potential difference across the grid). Solely a physical nozzle is also undesirable in such a plasma thruster as it will lead to wall erosion that reduces the lifespan of the thruster.

*2.4 Electrostatic Acceleration*

The final component of the hybrid plasma ion thruster is the electrode grids that facilitate the electrostatic acceleration of the ions. A beam neutralizer ejects the electrons separated from the plasma by the electrodes along the physical nozzle to maintain the charge parity of the spacecraft [18]. According to the Child-Langmuir relation, the voltage across infinite planar conductors (here the gridded electrodes) is proportional to the ion current density raised to the power ⅔ while the distance between the planes is inversely related to the square root of the ion current density. Typical ion thrusters operate at low densities to maximize the specific impulse, resulting in a low thrust. In the explored concept thruster, a high-density plasma requires to be accelerated, resulting in a high ion current density. Thus, the voltage needs to be high and the electrode gap minimized [19].

*2.5 Propellant*

For thrusters that use Ion Cyclotron Resonance Heating (ICRH) of plasma, it has been found by Otsuka et al. in [20] that Helium as a propellant outperforms Argon with respect to exhaust velocity, ion wall loss, and ion-neutral collisions. While Xenon is best suited for ion thrusters as it is the heaviest non-radioactive noble gas, it is expensive for utilization in high mass flow rate systems. Argon is thus very commonly used in ion thrusters as it is the next best non-reactive propellant which has a high mass and low ionization enthalpy which are optimum performance characteristics [21].

For the model that is simulated, Xenon, Argon, and Helium have been taken and the results compared. Xenon and Argon are the superior propellants for ion thrusters while lighter elements like Hydrogen and Helium are preferred for other plasma propulsion schemes.

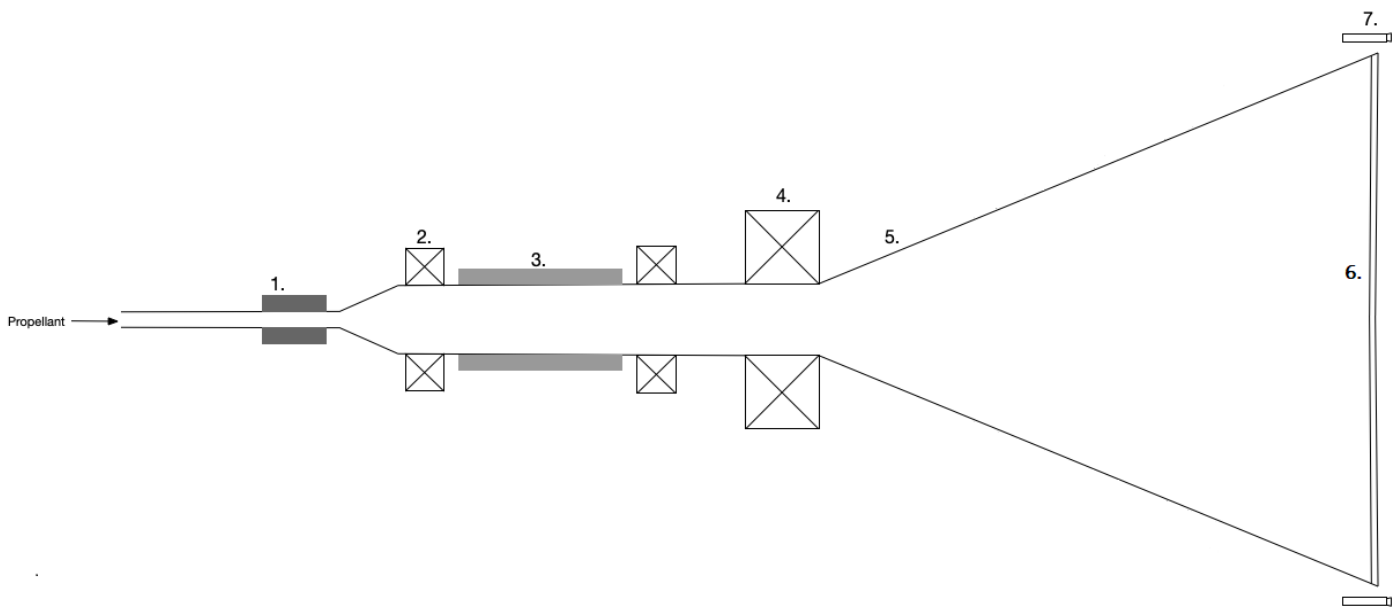

**Figure 1.** Schematic of explored thruster
1. Helicon Antenna; 2. Magnetic Mirror Coils; 3. ICRH antenna; 4. Magnetic Nozzle magnet; 5. Physical Nozzle; 6. Gridded electrodes for electrostatic acceleration; 7. Beam neutralizer

**3. Method**

*3.1 Description*

In the Particle in Cell (PIC) method, fluid and plasma systems are simulated as a collection of individual particles. The domain is divided into a mesh and at the vertices, or nodes, local plasma or fluid properties like density are calculated [22]. For the PIC simulation of the hybrid thruster in this paper, a Python3 code was written using the Spyder IDE. The code aimed to simulate the ICRH and Gridded Electrostatic Acceleration (GEA) and compare the exhaust velocities of the three propellants for different cases—only ICRH, only GEA, and with both systems on. Of the propellants that have been chosen—Helium, Argon, and Xenon (compared in Table 1)—the two heavier noble gases are commonly used in current ion thruster technology. The node spacing is well above the debye length for the three propellant plasmas, and thus a 2D hybrid electrostatic PIC (ES-PIC) has been used to simulate the system. The ion species have been simulated as separate particles, while electrons are considered as a fluid. The nodes have been distributed in a 220×53 grid, covering the upper half of the axisymmetric thruster (Figure 2). In order to account for the cylindrical geometry of the thruster in a 2D setup, individual node volumes have been adjusted based on the distance from the axis, with the lowest volume being at the nodes along the axis, and the highest volume at the boundary wall nodes. The volume is used while calculating the ion density, which is in turn used to ascertain the electrostatic potential at each node. Important parameters for the PIC code have been listed in Table 2.

The components of the simulated section are as follows:

1. Heating chamber - In this area, a cylinder of length 50 cm and radius of 1.5 cm is where the ion cyclotron resonance heating takes place;
2. Nozzle - Thermal expansion of the heated propellant occurs in the 60 cm-long conical space of the physical nozzle, the angle of which is 0.395 radians, thus reaching a maximum radius of 26.5 cm;
3. Electrode grid - Ions are accelerated through a 1 mm gap between two gridded electrodes each having a radius of 26.5 cm.

The initial ion velocities have been assigned based on the Maxwell-Boltzmann Velocity Distribution function of the species at a temperature of 5000K. The acceleration of the ions determined by the electric field is obtained from the local electric potential determined by using the Boltzmann electron relation:

$$n_e = n_o \exp(\frac{e(\phi_i - \phi_o)}{KT_e})  \quad (2)$$

This equation is solved for $\phi_i$ where $n_e$ is the local electron density at the node, $n_o$ is the average particle density of the system, e is the fundamental charge, $\phi_i$ is the local potential of the node, $\phi_o$ is the global potential inside the thruster (0 V), $K$ is the Boltzmann's constant, $T_e$ is the electron temperature and $n_i$ is the local ion density (between 0 - 0.5m of the simulation domain, this is a quasi-neutral plasma, $n_i = n_e$). Here, the $T_e$ is based on the findings of Sudit and Chen in [23] where the electron temperature measured in helicon plasmas averages 3.5 eV. This Quasi-Neutral Boltzmann Solver was utilized to cut down on computation time and is not a source for error as the final velocity depends almost entirely on the ICRH and GEA. For the ICRH, the energy change in the ions was based on the findings of Bering and colleagues in [24], where an average of ~17 eV was absorbed by individual ions. The maximum magnetic mirror field strength is 1000G.

After the ion cyclotron resonance heating, the ion particles exit the heating chamber. Above the nozzle throat, an electromagnet of 1700G is present, which acts as the magnetic nozzle. Similar to a de Laval nozzle used in non-plasma thrusters, the magnetic nozzle converts the thermal energy into directed kinetic energy along the axis. In the simulated model, the radial component of the ion particle velocity is reduced while the axial velocity component proportionally increases using the equation described below.

$$F_{||} = -\mu|\nabla B| \quad (3)$$

where $\mu = \frac{1}{2}m_i v_\perp^2 / B_{max}$, $B$ is the magnetic field, $m_i$ is the ion mass, $v_\perp$ is the ion velocity in the radial direction, and $B_{max}$ is the magnetic field strength at the nozzle throat. This simplified equation is used to calculate the axial velocity increase in the thruster, but for greater accuracy in analyzing electron and ion flow characteristics, more robust methods are required, as done by Ahedo and Merino in [25]. Since this study only deals with the velocity components of the simulated plasma, equation 3 is sufficient. The magnetic field strength of the magnetic nozzle and ICRH magnets along the thruster axis is shown in Figure 3.

In the nozzle domain, the propellant plasma is no longer in a quasi-neutral state as electrons are extracted by electrodes along the physical nozzle boundary. In order to calculate the node potential in this section of the thruster, the electron density ($n_e$) in equation 2 decreases from the axis to the nozzle boundary. The electron sheath potential in this section of the simulation domain follows a simple Debye sheath model [26] and is calculated with the formula:

$$\phi_i = \phi_{plate} \exp(\frac{-|x|}{\lambda_D}) + \phi_o \qquad (4)$$

where $\phi_{plate}$ is the potential at the nozzle walls, taken as 150V as in [27], x is the distance of the node from the nozzle wall, and $\lambda_D$ is the electron Debye length. While simulating solely the ICRH component of the thruster, the Boltzmann equation is used for calculating the node potentials in the nozzle area as there is no extraction of electrons in this permutation.

Prior to expulsion from the nozzle, the ions pass through the 1mm-wide electrode grid. For all three propellants, a voltage of 1000V across the electrode gap has been taken so that they can be better compared with respect to the thrust-specific impulse relation that has been derived in Appendix A. Moreover, this high potential is taken in accordance with the Child-Langmuir Limit as a high voltage is required for high currents of plasma flowing through a thin gap between two charged plates (here, the electrodes).

*3.2 Tables and Figures*

Table 1. Comparison of Mass and ionization enthalpies of the three propellants tested

| Propellant | Mass (u) | First Ionization Enthalpy (eV) | Initial Thermal Velocity (m/s) |
|---|---|---|---|
| Helium | 4 | 25 | 4,558 |
| Argon | 40 | 16 | 1,443 |
| Xenon | 131 | 12 | 796 |

Table 2. PIC Simulation Parameters and their values

| PIC Parameter | Value |
|---|---|
| Node spacing (dr & dz) | $5 \times 10^{-3}$ m |
| Domain size | 1.101 m × 0.265 m |
| Ion particle weight | $10^7$ |
| Time step (dt) | $5 \times 10^{-8}$ seconds |

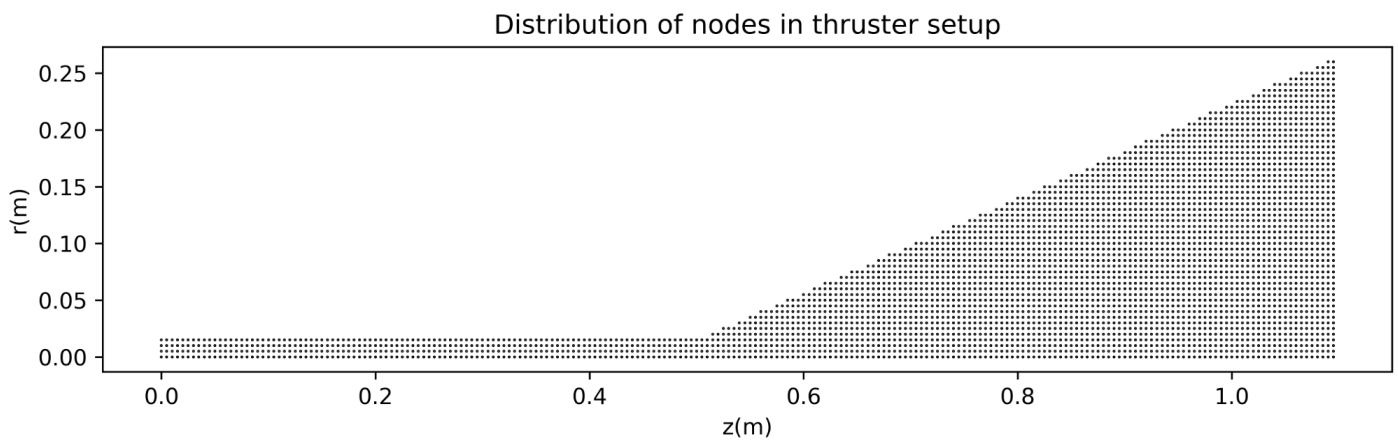

**Figure 2.** Distribution of nodes in thruster setup

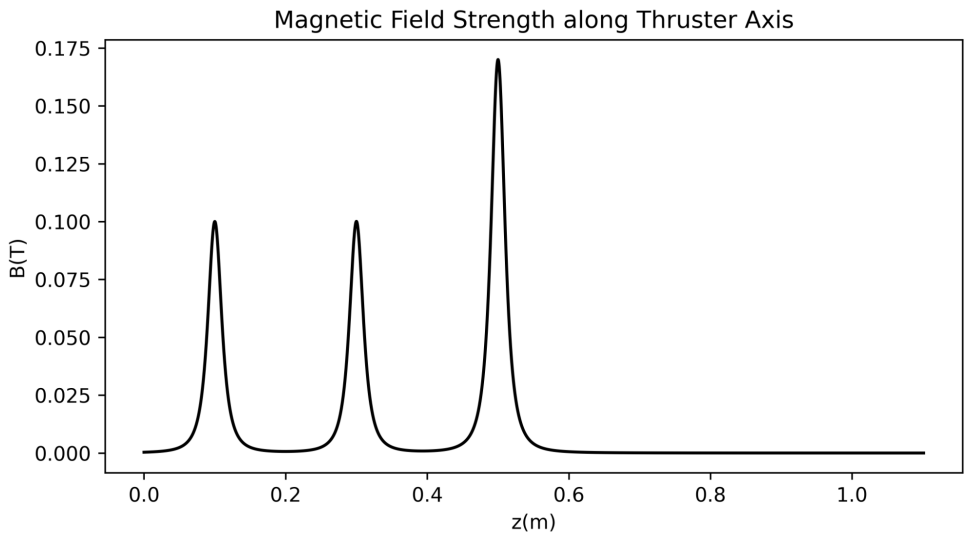

**Figure 3.** Magnetic Field Strength along Thruster Axis

## 4. Results

Based on the findings that have been listed in Table 3, there is a clear advantage in employing both ICRH and GEA in the thruster versus only either as the exhaust velocity is greater. With the combined system, Helium has the greatest velocity at 240 km/s, followed by Argon with 76 km/s and Xenon at 42 km/s. The difference in output between the only-ICRH and the combined system is significant, resulting in a ~151.2% increase. However, with this model of ICRH employed, the net velocity increase of both systems over solely the GEA is not substantial, averaging approximately 9.1% in all three cases. The specific impulse is calculated by dividing the exhaust velocity by the standard gravitational acceleration on Earth (9.8 m/s). While Helium has the highest specific impulse out of the three elements, this is not favorable for optimal thrust due to the inverse relationship between thrust and $I_{sp}$ in electric propulsion that has been derived in Appendix A. Moreover, taking production costs into consideration, Xenon requires the lowest amount of energy and is thus preferred in such plasma thrusters [21, 28]. The velocity ratio along the z and r axes also changes drastically, where $v_\perp/v_\parallel$ is approximately equal to one at initialization but drops to 0.016 for Xenon upon exiting the thruster. This is largely due to the axial velocity augmentation of the magnetic nozzle although the electrostatic acceleration component, which is responsible for the majority of the particle's velocity and only induces motion along the z-axis, also contributes to this low ratio.

**Table 3.** Exhaust velocity and specific impulse recordings for Helium, Argon, and Xenon

| Propellant | Component Simulated | Exhaust Velocity (m/s) | Specific Impulse (s) |
|---|---|---|---|
| | ICRH | 90,659 | 9,251 |
| Helium | GEA | 221,827 | 22,635 |
| | ICRH & GEA | 239,638 | 24,453 |
| | ICRH | 31,876 | 3,253 |
| Argon | GEA | 68,577 | 6,997 |
| | ICRH & GEA | 75,623 | 7,717 |
| | ICRH | 16,557 | 1,689 |
| Xenon | GEA | 38,322 | 3,910 |
| | ICRH & GEA | 41,746 | 4,260 |

## 5. Discussion

The combination of electrostatic acceleration and ion cyclotron resonance heating results in a system that is the middle ground between high specific impulse and high thrust, particularly with heavy propellants contrasted to existing ion thrusters which have high $I_{sp}$ but low thrust rates, and chemical thrusters which produce large thrust values at a low $I_{sp}$. In line with existing ion thrusters, Xenon is the best propellant due to its low first ionization enthalpy and high density. There is a relatively low increase in the exhaust velocity due to the ICRH versus the GEA, and, as expected, ion thrusters are better suited for low payload missions, whereas the hybrid thruster could be used for large-payload, high-thrust spacecraft. A Xenon or Argon propelled system as simulated can be optimized by improving the ICRH component to maximize thrust.

Among the three propellants studied, Helium has the highest specific impulse. However, since the Gridded Electrostatic Acceleration driven by electric power dominates, Xenon, with a lower $I_{sp}$ avails a higher thrust. Hence, among Xenon and Helium electric thrusters with the same power input, Xenon is the optimal propellant. The inverse relation between specific impulse and thrust in electric propulsion has been derived in Appendix A. Xenon is also generally preferred as it does not condense on the nozzle and accelerator grids, and is easy to store due to its high boiling point compared to Helium (Goebel & Katz, 2008, p. 1-2).

The power requirement of this system is immense in comparison to that of ion thrusters due to the combined cost of the helicon and ICR antennas, and the electrode grid. Such a thruster cannot be powered by solar panels, especially since such a system is advantageous for deep space missions that are typically further from the Sun. Thus, nuclear energy is a possible energy source, with fission-derived energy being the most practical with current technology [29]. Progress in fusion technology could hold the key for future high-power electric and plasma propulsion systems, as the fuel is cheaper, more abundant than fission fuel, and does not pose any radiation-induced challenges. However, if advances in technology render a Direct Fusion Drive to be feasible, it would be preferred over fusion-electric propulsion as the former expels the products of a nuclear fusion reaction would have a far greater energy efficiency than converting fusion power to electric power for this model [30].

Another propulsion system that has gained traction over the last few decades is that of a magnetic nozzle with ICR-heated plasmas, popularized by the VASIMR. Here, the magnetic nozzle, which changes its field strength based on the flow rate, efficiently converts the thermal energy into kinetic energy along the direction of mass flow. Essentially, the additional thrust that the system discussed in this paper can be compared to magnetic nozzles in such plasma thrusters. While the latter system has its advantages, work needs to be done to study the viability of electrode grids on plasma rockets.

The specific impulse of the thruster explored in this paper stands at 7,717 s for Argon, which is significantly higher than those of existing plasma thrusters like the VASIMR [5], and the High Power Helicon Thruster developed by Ziemba et. al [31], which afford 5400 s and 1500 s respectively with Argon. The model's $I_{sp}$ with Xenon (4260 s) is comparable with NASA's NEXT ion thruster [6] which has an experimental specific impulse of 4190 s.

A significant limitation of the system described in this paper, however, is the increased grid erosion as a result of the ionized gas that enters the grid at higher velocities than in ion thrusters. While increased erosion can be attenuated with a decelerator grid [32], further advancements still need to be made in grid technology to increase the lifespan of such thrusters.

While in the results discussed in this paper the contribution of the ICRH is far lower than the GEA, an optimized ICRH system could potentially play a more significant role in the final exhaust velocity, in which case medium-density propellants like Argon would be most suitable.

## 6. Conclusions

Conventional rocket propulsion technology is not suited for large payload deep space missions. A hybrid plasma-ion thruster maximizes thrust while maintaining a high specific impulse can facilitate quicker and more efficient space travel. In this paper, such a hybrid thruster concept is explored which consists of a helicon antenna for plasma generation, plasma heating by ion cyclotron resonance, conversion of thermal to kinetic energy through a magnetic nozzle, and electrostatic acceleration of ions through gridded electrodes.

Through a hybrid particle-in-cell simulation, it was found that the combination of gridded electrostatic acceleration and ion cyclotron resonance heating results in an increased exhaust velocity and thrust that exceeds either of the two systems in isolation. The average exhaust velocities for Helium, Argon, and Xenon were recorded to be 240 km/s, 76 km/s, and 42 km/s respectively. As the velocity increase due to electrostatic acceleration dominates in this system, heavier elements like Xenon and Argon are preferred over lighter elements like Helium as is the case with conventional ion thrusters. While the preliminary simulations indicate that the advantages of the hybrid thruster to ion thrusters seem small, more work is needed to better describe the optimum conditions of the separate stages of the thruster, particularly the ion cyclotron resonance heating in order to maximize both the thrust and specific impulse for large payload space missions.

## 7. Figure Captions

**Figure 1.** Schematic of explored thruster

1. Helicon Antenna; 2. Magnetic Mirror Coils; 3. ICRH antenna; 4. Magnetic Nozzle magnet; 5. Physical Nozzle; 6. Anodes for extracting electrons; 7. Gridded electrodes for electrostatic acceleration; 8. Beam neutralizer

**Figure 2.** Distribution of nodes in thruster setup

**Figure 3.** Magnetic Field Strength along Thruster Axis

**Supplementary Materials:**
[dataset] Mishra, Ishaan (2021), "Thruster Particle-in-Cell Simulation", Mendeley Data, V3, DOI: 10.17632/2r88w25743.3

**Acknowledgments:** I wish to show my appreciation to Geetha Sen of the Liquid Propulsion System Centre, ISRO, and Professor Sanjeev Varshney at the Institute of Plasma Research for their academic support. I would also like to thank Professor John Foster at the University of Michigan and Professor Ken Hara at Stanford University for directing me to the right resources for writing particle-in-cell simulations.

**Funding:** This research received no external funding.

**Conflicts of Interest:** The author declares no conflict of interest.

## Appendix A

In electric propulsion, where electric energy is converted into mechanical energy, a clear relationship can be derived between the thrust ($F$) and specific impulse ($I_{sp}$) [33]. This is relevant to the model explored in this paper as the majority of the thrust is obtained from the gridded electrostatic thruster. This relation can be derived from the following three equations:

$$I_{sp} = \frac{F}{\dot{m}g} \tag{A1}$$

where $\dot{m}$ is the mass flow rate of the propellant, $v$ is the exhaust velocity and $g$ is the standard acceleration due to gravity on Earth.

$$F = \dot{m}v$$

$$P_e = \frac{1}{2}\dot{m}v^2$$

Where $P_e$ is the electric power supplied. Manipulating these equations,

$$I_{sp} = \frac{v}{g}$$

$$P_e = \frac{1}{2}gFI_{sp}$$

$$F = \frac{2P_e}{I_{sp} \times g} \tag{A2}$$

Equation A2 shows that for constant electric power there is an inverse relationship in ion thrusters, between the $I_{sp}$ and the thrust.